# Self-consistent solution of Eliashberg equations for metal hydride superconductors

**Tomas J. Escamilla and Chumin Wang***

Instituto de investigaciones en Materiales, Universidad Nacional Autónoma de México, Mexico City, MEXICO

*E-mail: chumin@unam.mx

**Abstract.** In recent years, the quest for high critical-temperature superconductors has increasingly focused on metal and molecular hydrides, which have demonstrated potential for superconductivity at or near room temperature under extremely high pressures. Such hydrides were first proposed by N. W. Ashcroft in 1968, because hydrogen-rich materials possess elevated vibrational frequencies due to the low atomic mass of hydrogen. This article presents a self-consistent solution to the Eliashberg equations for analysing superconductivity in hydrides, contrasting with the commonly used McMillan-Allen-Dynes parameterized formula. We also analyse effects of the electron-phonon spectral function and the broadening parameters arising from phonon lifetime and sample imperfections on the superconducting critical temperature. Finally, both theoretical approaches are applied to a typical metal hydride superconductor, and the reliability of self-consistent solutions is validated against the experimentally measured critical temperature.

## 1. Introduction

Superconductivity was found in 1911 by H. K. Onnes, who observed it in pure mercury at 4.15 K. For several decades, liquid helium was required to achieve the low temperatures for its study [1]. In 1986, the discovery of ceramic superconductors enabled the use of liquid nitrogen as a coolant [2-3]. Superconductivity at room temperature has been a longstanding challenge and a dream for generations of condensed matter physicists. In 2015, metallic $H_3S$ was identified as the first hydride superconductor with a critical temperature ($T_c$) of 203 K under a pressure of 200 GPa [4]. Stimulated by this experimental breakthrough, other high-pressure hydride superconductors were found, such as $YH_{10}$ [5], $LaH_{10}$ [6], $ThH_{10}$ [7], $YH_6$ [8], and $CaH_6$ [9]. These compounds share a common high density of hydrogen atoms, as suggested by N. W. Ashcroft in 1968 [10].

On the theoretical side, the first microscopic theory of superconductivity was developed in 1957 by J. Bardeen, L. Cooper, and J. R. Schrieffer (BCS) [11]. Three years later, G. M. Eliashberg [12] refined this theory by explicitly incorporating phonon dynamics and retardation effects, which gave rise to an electron-phonon spectral function $\alpha^2 F(\omega)$ that can be experimentally determined via tunnelling spectroscopy [13] and angle-resolved photoemission spectroscopy (ARPES) [14]. In 1968, W. L. McMillan found an analytical solution to the linearized Eliashberg equations [15], which was further improved by J. V. Allen and R. C. Dynes through fitting the $T_c$ of many superconductors known at that time [16].

Nowadays, the McMillan-Allen-Dynes formula is widely used to calculate the superconducting $T_c$. For example, both the critical temperature and external pressure required for the first hydride superconductor H$_3$S were successfully predicted by the mentioned McMillan-Allen-Dynes formula, using $\alpha^2 F(\omega)$ obtained from the Density Functional Theory (DFT) [17]. In the present article, we present a self-consistent solution of the Eliashberg equations that determines the $T_c$ by imposing the condition of vanishing superconducting gaps in Matsubara space. As an example, we apply this solution to calcium hydride CaH$_6$, whose $\alpha^2 F(\omega)$ is computed using the Quantum ESPRESSO software within the Density Functional Perturbation Theory (DFPT) framework [18]. This self-consistent solution of $T_c$ is contrasted with that obtained from the McMillan-Allen-Dynes formula, as well as with the experimental data. In the next section, we will introduce the Eliashberg theory based on the Green's function and Matsubara imaginary frequencies.

## 2. Short derivation of Eliashberg equations

In 1958, A. B. Migdal created a formalism for electron-phonon interaction in metals using Green's functions and imaginary time [19]. Two years later, G. Eliashberg extended this formalism to the superconducting state, explicitly considering phonons as mediators of electron pairing [12]. In the same year, Y. Nambu reformulated Migdal's formalism using 2×2 matrix notation for the Green's function ($\hat{G}$) in terms of the Matsubara imaginary time ($\tau = it$) given by [20]

$$\hat{G}(\vec{k},\tau) = -<\hat{T}_\tau \hat{\Psi}_{\vec{k}}(\tau)\hat{\Psi}_{\vec{k}}^\dagger(0)> = -\begin{pmatrix} <\hat{T}_\tau \hat{c}_{\vec{k},\uparrow}(\tau)\hat{c}_{\vec{k},\uparrow}^\dagger(0)> & <\hat{T}_\tau \hat{c}_{\vec{k},\uparrow}(\tau)\hat{c}_{-\vec{k},\downarrow}(0)> \\ <\hat{T}_\tau \hat{c}_{-\vec{k},\downarrow}^\dagger(\tau)\hat{c}_{\vec{k},\uparrow}^\dagger(0)> & <\hat{T}_\tau \hat{c}_{-\vec{k},\downarrow}^\dagger(\tau)\hat{c}_{-\vec{k},\downarrow}(0)> \end{pmatrix}, \quad (1)$$

where $\hat{T}_\tau$ is the Wick operator that arranges the imaginary time, $\hat{\Psi}_{\vec{k}}^\dagger(\tau) = \left(\hat{c}_{\vec{k},\uparrow}^\dagger(\tau), \hat{c}_{-\vec{k},\downarrow}(\tau)\right)$ is a two-component field operator, and $\hat{c}_{\vec{k},\sigma}^\dagger$ ($\hat{c}_{\vec{k},\sigma}$) is the creation (annihilation) operator of single electron in the Bloch state $(\vec{k},\sigma)$ with spin $\sigma \in \{\uparrow,\downarrow\}$. Notice that off-diagonal elements in Eq. (1) are associated with Cooper pairs, while the diagonal ones describe the normal state.

In atomic units with $k_B = \hbar = 1$, the Fourier transform of Green's function in Eq. (1) is [21,22]

$$\hat{G}(\vec{k},\tau) = \frac{1}{\beta} \sum_{n=-\infty}^{\infty} e^{-i\omega_n \tau} \hat{G}(\vec{k},i\omega_n), \quad (2)$$

where $\beta = 1/T$, $\omega_n = (2n+1)\pi/\beta$ is fermion Matsubara frequency with *n* an integer number, and

$$\hat{G}(\vec{k},i\omega_n) = \int_0^\beta d\tau\, e^{i\omega_n \tau} \hat{G}(\vec{k},\tau). \quad (3)$$

The Dyson equation for $\hat{G}(\vec{k},i\omega_n)$ can be written as [23]

$$\hat{G}^{-1}(\vec{k},i\omega_n) = \hat{G}_0^{-1}(\vec{k},i\omega_n) - \hat{\Sigma}(\vec{k},i\omega_n) \quad (4)$$

where $\hat{G}_0(\vec{k},i\omega_n) = [i\omega_n \sigma_0 - \varepsilon(\vec{k})\sigma_3]^{-1}$ is the Green's function for electrons in normal state with $\varepsilon(\vec{k})$ the electronic dispersion relation and $\hat{\Sigma}(\vec{k},i\omega_n)$ is the self-energy written as a 2×2 matrix in accordance with Eq. (1). This self-energy can be expressed as [24]

$$\hat{\Sigma}(\vec{k},i\omega_n) = i\omega_n[1 - z(\vec{k},i\omega_n)]\sigma_0 + \chi(\vec{k},i\omega_n)\sigma_3 + \phi(\vec{k},i\omega_n)\sigma_1 + \tilde{\phi}(\vec{k},i\omega_n)\sigma_2, \quad (5)$$

where $z(\vec{k},i\omega_n)$ is the mass renormalization function, $\chi(\vec{k},i\omega_n)$ is the energy shift, $\phi(\vec{k},i\omega_n)$ and $\tilde{\phi}(\vec{k},i\omega_n)$ are the order parameters, and

$$\sigma_0 = \begin{pmatrix} 1 & 0 \\ 0 & 1 \end{pmatrix}, \ \sigma_1 = \begin{pmatrix} 0 & 1 \\ 1 & 0 \end{pmatrix}, \ \sigma_2 = \begin{pmatrix} 0 & -i \\ i & 0 \end{pmatrix} \ \text{and} \ \sigma_3 = \begin{pmatrix} 1 & 0 \\ 0 & -1 \end{pmatrix}. \tag{6}$$

Substituting Eq. (5) into Eq. (4), we obtain

$$G(\vec{k},i\omega_n) = \frac{1}{\det(G^{-1})}\left[i\omega_n z(\vec{k},i\omega_n)\sigma_0 + \left(\varepsilon(\vec{k}) + \chi(\vec{k},i\omega_n)\right)\sigma_3 + \phi(\vec{k},i\omega_n)\sigma_1 + \tilde{\phi}(\vec{k},i\omega_n)\sigma_2\right], \tag{7}$$

where $\phi$ and $\tilde{\phi}$ are off-diagonal elements of the Green's function and

$$\det(G^{-1}) = (i\omega_n z)^2 - [\varepsilon(\vec{k}) + \chi]^2 - \phi^2 - \tilde{\phi}^2. \tag{8}$$

Performing an analytical continuation $i\omega_n \to E_{\vec{k}} + i\delta$, the poles of Green's function $G(\vec{k},i\omega_n)$ in Eq. (7) obtained from $\det(G^{-1}) = 0$ are located at [21]

$$E_{\vec{k}} = \sqrt{\frac{[\varepsilon(\vec{k}) + \chi]^2}{z^2} + \frac{\phi^2 + \tilde{\phi}^2}{z^2}} = \sqrt{\frac{[\varepsilon(\vec{k}) + \chi]^2}{z^2} + |\Delta|^2}, \tag{9}$$

where $\Delta$ is the superconducting gap given by

$$\Delta(\vec{k},i\omega_n) = \frac{\phi(\vec{k},i\omega_n) - i\tilde{\phi}(\vec{k},i\omega_n)}{z(\vec{k},i\omega_n)}. \tag{10}$$

Following the Migdal's theorem, self-energy in Eq. (4) can be written as [19,24]

$$\hat{\Sigma}(\vec{k},i\omega_n) = \hat{\Sigma}_{ep}(\vec{k},i\omega_n) + \hat{\Sigma}_c(\vec{k},i\omega_n), \tag{11}$$

where the electron-phonon contribution is

$$\hat{\Sigma}_{ep}(\vec{k},i\omega_n) \simeq -\frac{1}{\beta}\sum_{\vec{k}',n'}\sigma_3 \hat{G}(\vec{k}',i\omega_{n'})\sigma_3 \sum_\nu |g^\nu_{k,k'}|^2 D_\nu(\vec{k}-\vec{k}',i\omega_n - i\omega_{n'}) \tag{12}$$

and the Coulomb contribution is

$$\Sigma_c(\vec{k},i\omega_n) \simeq -\frac{1}{\beta}\sum_{\vec{k}',n'}\sigma_3 \hat{G}^{od}(\vec{k}',i\omega_{n'})\sigma_3 \tilde{V}(\vec{k}-\vec{k}'), \tag{13}$$

being $g^\nu_{k,k'}$ the electron-phonon interaction matrix and

$$D_\nu(\vec{k}-\vec{k}',i\omega_n - i\omega_{n'}) \approx D^{(0)}_\nu(\vec{k}-\vec{k}',i\omega_n - i\omega_{n'}) = \frac{2\omega_\nu(\vec{k}-\vec{k}')}{(i\omega_n - i\omega_{n'})^2 - [\omega_\nu(\vec{k}-\vec{k}')]^2} \tag{14}$$

is the correlation function for harmonic phonons with wave vector $\vec{q} = \vec{k} - \vec{k}'$, angular frequency $\omega_\nu(\vec{q})$ of phonon branch $\nu$, and bosonic Matsubara frequency $\Omega_m = \omega_n - \omega_{n'} = 2m\pi/\beta$ [21]. In Eq. (13) $\hat{G}^{od}(\vec{k}',i\omega_{n'}) = (\phi\sigma_1 + \tilde{\phi}\sigma_2)/\det(\hat{G}^{-1})$ is $\hat{G}(\vec{k}',i\omega_{n'})$ with null diagonal elements and $\tilde{V}(\vec{k}-\vec{k}')$ is the screened Coulomb potential.

Given that $\sigma_0$, $\sigma_1$, $\sigma_2$ and $\sigma_3$ are linearly independent, substituting Eqs. (5) and (7) into Eq. (11), we obtain the famous Eliashberg equations [12,25]

$$\begin{cases} \omega_n[1-z(\vec{k},i\omega_n)] = -\dfrac{1}{\beta}\sum_{\vec{k}',n'}\dfrac{\omega_{n'}z(\vec{k}',i\omega_{n'})}{\det(G^{-1})}\sum_{\nu}|g^{\nu}_{k,k'}|^2 D_{\nu}(\vec{k}-\vec{k}',i\omega_n-i\omega_{n'}) \\ \chi(\vec{k},i\omega_n) = -\dfrac{1}{\beta}\sum_{\vec{k}',n'}\dfrac{\varepsilon(\vec{k}')+\chi(\vec{k}',i\omega_{n'})}{\det(G^{-1})}\sum_{\nu}|g^{\nu}_{k,k'}|^2 D_{\nu}(\vec{k}-\vec{k}',i\omega_n-i\omega_{n'}) \\ \Phi(\vec{k},i\omega_n) = \dfrac{1}{\beta}\sum_{\vec{k}',n'}\dfrac{\Phi(\vec{k}',i\omega_{n'})}{\det(G^{-1})}\left[\sum_{\nu}|g^{\nu}_{k,k'}|^2 D_{\nu}(\vec{k}-\vec{k}',i\omega_n-i\omega_{n'})+\tilde{V}(\vec{k}-\vec{k}')\right] \end{cases} \quad (15)$$

where $\Phi$ represents $\phi$ or $\tilde{\phi}$. In the following, we choose the gauge where $\tilde{\phi}=0$. Thus, Eq. (10) becomes $\Delta(\vec{k},i\omega_n)=\phi(\vec{k},i\omega_n)/z(\vec{k},i\omega_n)$, which has a nontrivial solution $\left(\Delta(\vec{k},i\omega_n)\neq 0\right)$ from Eqs. (15) for each temperature and then, the highest temperature corresponds to $T_c$.

On the other hand, the Eliashberg spectral function $\alpha^2 F(\omega)$ averaged over the Fermi surface can be written as [25]

$$\alpha^2 F(\omega) = \dfrac{1}{N(0)}\sum_{\vec{k},\vec{k}'}\sum_{\nu}|g^{\nu}_{k,k'}|^2 \delta\left(\varepsilon(\vec{k})\right)\delta\left(\varepsilon(\vec{k}')\right)\delta\left(\omega-\omega_{\nu}(\vec{k}-\vec{k}')\right). \quad (16)$$

Hence, the electron-phonon coupling parameter is given by:

$$\lambda(i\omega_n - i\omega_{n'}) = \int_0^{\infty}\dfrac{2\omega\,\alpha^2 F(\omega)}{(\omega_n-\omega_{n'})^2+\omega^2}d\omega. \quad (17)$$

Since superconducting pairing predominantly occurs within a narrow energy window around the Fermi level, it is common practice to simplify the Eliashberg equations by restricting them to electronic states close to the Fermi energy, which leads to $\chi(\vec{k},i\omega_n)=0$ [24]. Now, substituting Eqs. (14), (16) and (17) in Eqs. (15), we obtain isotropic Eliashberg equations given by

$$\begin{cases} z(i\omega_n) = \left\langle z(\vec{k},i\omega_n)\right\rangle_{\varepsilon(\vec{k})=0} = 1+\dfrac{\pi}{\omega_n\beta}\sum_{n'}\dfrac{\omega_{n'}}{\sqrt{(\omega_{n'})^2+\Delta^2(i\omega_{n'})}}\lambda(i\omega_n-i\omega_{n'}) \\ \Delta(i\omega_n) = \dfrac{\left\langle \phi(\vec{k},i\omega_n)\right\rangle_{\varepsilon(\vec{k})=0}}{z(i\omega_n)} = \dfrac{\pi}{z(i\omega_n)\beta}\sum_{n'}\dfrac{\Delta(i\omega_{n'})}{\sqrt{(\omega_{n'})^2+\Delta^2(i\omega_{n'})}}\left[\lambda(i\omega_n-i\omega_{n'})-\mu^*\right] \end{cases}. \quad (18)$$

where $\mu^* = N(0)\left\langle\tilde{V}(\vec{k}-\vec{k}')\right\rangle_{\varepsilon(\vec{k}),\varepsilon(\vec{k}')=0}$ is the dimensionless Coulomb interaction parameter.

## 3. Results

First-principles calculations in this article were performed in two stages: (i) Geometry optimization using CASTEP within *Materials Studio* framework [26] and (ii) Calculation of $\alpha^2 F(\omega)$ via *Quantum ESPRESSO* (QE) package [18]. Both stages utilise the Perdew–Burke–Ernzerhof exchange-correlation functional [27] with CASTEP norm-conserving [28] and QE projector augmented-wave (PAW) [29] pseudopotentials. To numerically evaluate the delta functions in Eq. (16), we introduce phonon ($\eta_{ph}\to 0$) and electron-phonon interaction ($\eta_{el-ph}\to 0$) broadening parameters to include structural disorder and finite phonon lifetime effects [30]. Table 1 summarises key computational parameters used in this study.

Figure 1 shows the Eliashberg spectral function $\alpha^2 F(\omega)$ for $CaH_6$ at 200 GPa with a cubic *Im-3m* structure illustrated in the inset for $\mu^*=0.1$, $\eta_{el-ph}=0.03$ Ry (red circles) and $\eta_{el-ph}=0.006$ Ry (violet squares). Notice that the main peaks of $\alpha^2 F(\omega)$ confirm those reported for $CaH_6$ at 150 GPa [31] and at 200 GPa [32,33], as well as the appearance of sharp peaks as $\eta_{el-ph}$ diminishes.

**Table 1.** Summary of the parameters used in numerical calculations

| Physical quantity | Geometry optimization (DFT by CASTEP) | Spectral function $\alpha^2 F(\omega)$ (DFPT by QE) |
|---|---|---|
| Total energy tolerance | $5.0\times 10^{-6}$ eV/atom | $10^{-15}$ Ry |
| Maximum force tolerance | 0.01 eV/Å | |
| Maximum stress tolerance | 0.02 GPa | |
| Maximum displacement tolerance | 0.0005 Å | |
| Plane-wave energy cutoff | 850 eV | 77 Ry |
| Broadening method | | Marzari-Vanderbilt |
| Phononic broadening $\eta_{ph}$ | | 0.18 THz |
| Electron-phonon interaction broadening $\eta_{el-ph}$ | | 0.006-0.045 Ry |
| Separation or number of k-point for electrons | 0.07 Å$^{-1}$ | 90×90×90 |
| Number of q-point for phonons | | 9×9×9 |

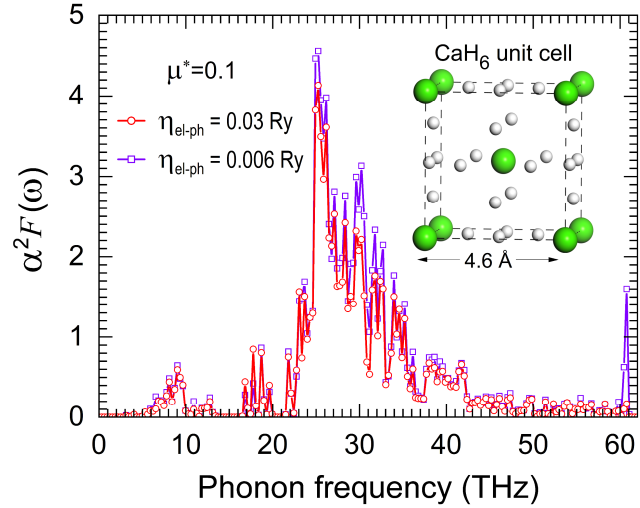

**Figure 1.** Eliashberg spectral function of CaH$_6$, whose unit cell is shown in the inset, for $\mu^* = 0.1$ with broadening parameter $\eta_{el-ph} = 0.03$ Ry (red circles) and $\eta_{el-ph} = 0.006$ Ry (violet squares).

The superconducting gap $\Delta(i\omega_n)$ obtained from Eqs. (18) is plotted in Figure 2 as a function of Matsubara frequencies ($\omega_n$) for $\mu^* = 0.1$, $\eta_{el-ph} = 0.021$ Ry and several temperatures from 25 K to 223 K, where $\omega_D$ is the Debye frequency. These $\Delta(i\omega_n)$ were obtained by self-consistently solving Eqs. (18), which consists of assuming an ansatz of $\{\Delta(i\omega_n) | n = 1, \cdots, \infty\}$, calculating the new $\{\Delta(i\omega_n)\}$ via Eqs. (18), and repeating this procedure until the difference of two consecutive solutions is smaller than a tolerance value. Observe that the magnitude of self-consistently obtained $\Delta(i\omega_n)$ reduces with increasing temperature. Hence, the critical temperature $T_c$ can be defined as the minimal temperature at which the superconducting gap $\Delta(i\omega_n)$ vanishes for all Matsubara frequencies.

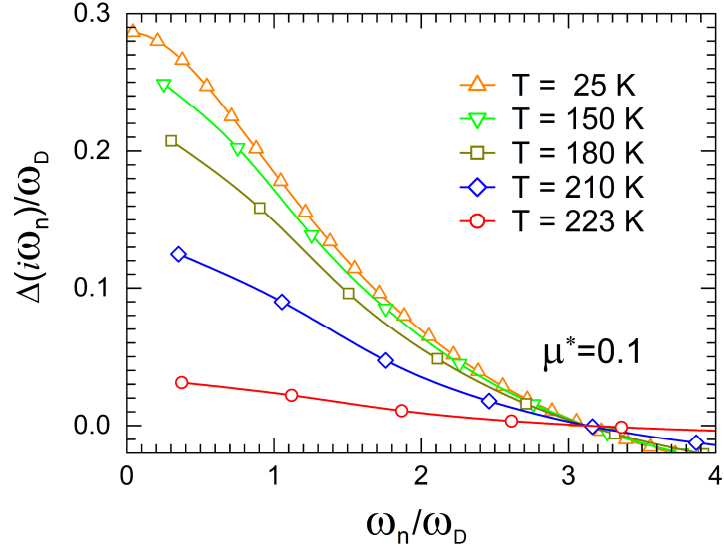

**Figure 2.** Superconducting gap (Δ) versus Matsubara frequencies ($\omega_n$) for $\eta_{el-ph}$= 0.021 Ry, $\mu^*$= 0.1, and temperature of 25 K (orange up triangles), 150 K (green down tringles), 180 K (gold squares), 210 K (blue rhombus), and 223 K (red circles).

Figure 3 shows the variation of $T_c$ as a function of the electron-phonon interaction broadening parameter ($\eta_{el-ph}$), obtained from the self-consistent solution (red circles) in comparison with those (gold triangles) of the McMillan-Allen–Dynes formula [16], both contrasting to the experimental data (blue dashed line) [9]. Note that the self-consistent solutions generally overestimate $T_c$ in comparison with the McMillan-Allen–Dynes formula [34], and in this case they are closer to the experimental data.

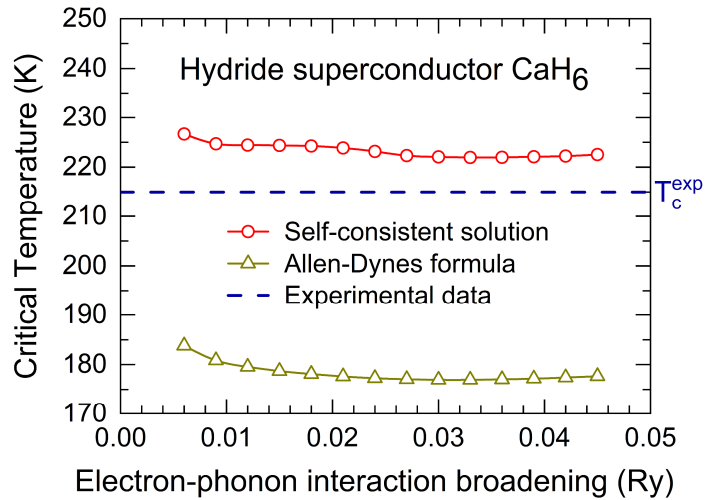

**Figure 3.** Superconducting critical temperatures versus $\eta_{el-ph}$ obtained from the self-consistent solution (red circles) and from the McMillan-Allen-Dyne formula (gold triangles), in comparison with experimental data (blue dashed line) [9].

## 4. Conclusions

In this article, we have presented a self-consistent solution for the critical temperature $T_C$ to the isotropic Eliashberg equations for s-wave superconductors by requiring the null Matsubara-space superconducting gaps, as schematically illustrated in Figure 2. This solution has the advantage of being simple and does not require experimental input. We further applied this self-consistent method to calcium hydride ($CaH_6$) under a pressure of 200 GPa, performing an ab-initio DFT plus DFPT calculations to obtain the Eliashberg spectral function $\alpha^2 F(\omega)$ for different broadening parameters. Using this spectral function, we self-consistently solved Eqs. (18) to determine $T_C$, which exhibits better agreement with the experimental data compared to those obtained from the semi-empirical McMillan-Allen-Dynes formula.

In general, the self-consistent solution of the Eliashberg equations provides a reliable approach to predict $T_C$ of conventional superconductors, including hydrides under high external pressures. However, such solutions often overestimate $T_C$ and they sensitively depend on the electron-phonon interaction broadening parameter, whose minimum reliable value is determined by the numerical calculation accuracy, particularly by the density of the ***k***- and ***q***-meshes respectively in the first electronic and phononic Brillouin zones. The present study can be extended to analyse anisotropic superconductors via Eqs. (15), especially when the Fermi energy lies close to van Hove singularities.

## Acknowledgements

This work has been partially supported by the Secretaría de Ciencias, Humanidades, Tecnología e Innovación de México (SECIHTI) through grant CF-2023-I-830 and by the Universidad Nacional Autónoma de México (UNAM) through project PAPIIT-IN110823. Computations were performed at Miztli under the grant of LANCAD-UNAM-DGTIC-039. The technical assistances of Alejandro Pompa, Oscar Luna, Cain Gonzalez, Silvia E. Frausto and Yolanda Flores are fully appreciated. T.J.E. acknowledges the fellowship for Master's Degree study from SECIHTI.

## References

[1] C. P. Poole, H. A. Farach, R. J. Creswick, and R. Prozorov, *Superconductivity* (Elsevier, Amsterdam, 2007) p. 24.

[2] J. G. Bednorz and K. A. Muller, Possible high Tc superconductivity in the Ba-La-Cu-O system, *Z. Phys. B-Condensed Matter* **64**, 189-193 (1986).

[3] M. K. Wu, J. R. Ashburn, C. J. Torng, P. H. Hor, R. L. Meng, L. Gao, Z. J. Huang, Y. Q. Wang, and C. W. Chu, Superconductivity at 93K in a new mixed-phase Y-Ba-Cu-O compound system at ambient pressure, *Phys. Rev. Lett.* **58**, 908-910 (1987).

[4] A. P. Drozdov, M. I. Eremets, I. A. Troyan, V. Ksenofontov, and S. I. Shylin, Conventional superconductivity at 203 Kelvin at high pressures in the sulfur hydride system, *Nature* **525**, 75-76 (2015).

[5] F. Peng, Y. Sun, C. J. Pickard, R. J. Needs, Q. Wu, and Y. Ma, Hydrogen clathrate structures in rare earth hydrides at high pressure: possible route to room-temperature superconductivity, *Phys. Rev. Lett.* **119**, 107001 (6pp) (2017).

[6] M. Somayazulu, M. Ahart, A. K. Mishra, Z. M. Geballe, M. Baldini, Y. Meng, V. V. Struzhkin, and R. J. Hemley, Evidence for superconductivity above 260 K in lanthanum superhydride at megabar pressure, *Phys. Rev. Lett.* **122**, 027001 (6pp) (2019).

[7] D. V. Semenok, A. G. Kvashnin, A. G. Ivanova, V. Svitlyk, V. Yu. Fominski, A. V. Sadakov, O. A. Sobolevskiy, V. M. Pudalov, I. A. Troyan, and A. R. Oganov, Superconductivity at 161 K in thorium hydride $ThH_{10}$: Synthesis and properties, *Materials Today* **33**, 36-44 (2020).

[8] I. A. Troyan, D. V. Semenok, A. G. Kvashnin, A. G. Ivanova, V. B. Prakapenka, V. S. Minkov, S. P. Besedin, M. A. Kuzovnikov, S. Mozaffari, A. R. Oganov, and M. I. Eremets, Anomalous high-temperature superconductivity in $YH_6$ and $YH_9$ under high pressure, *Adv. Mater.* **33**, 2006832 (10pp) (2021).

[9] L. Ma, K. Wang, Y. Xie, X. Yang, Y. Wang, M. Zhou, H. Liu, X. Yu, Y. Zhao, H. Wang, G. Liu, and Y. Ma, High-temperature superconducting phase in clathrate calcium hydride CaH6 up to 215 K at a pressure of 172 GPa, *Phys. Rev. Lett.* **128**, 167001 (6pp) (2022).

[10] N. W. Ashcroft, Metallic hydrogen a high-temperature superconductor?, *Phys. Rev. Lett.* **26**, 1748-1749 (1968).

[11] J. Bardeen, L. N. Cooper, and J. R. Schrieffer, Theory of superconductivity, *Phys. Rev.* **108**, 1175-1204 (1957).

[12] G. M. Eliashberg, Interactions between electrons and lattice vibrations in a superconductor, *Soviet Phys. JEPT* **11**(3), 696-702 (1960).

[13] L. Y. L. Shen, Tunneling into a high-Tc superconductor-Nb$_3$Sn, *Phys. Rev. Lett.* **29**, 1082-1085 (1972).

[14] W. Meevasana, F. Baumberger, K. Tanaka, F. Schmitt, W. R. Dunkel, D. H. Lu, S.-K. Mo, H. Eisaki, and Z.-X. Shen, Extracting the spectral function of the cuprates by a full two-dimensional analysis: Angle-resolved photoemission spectra of Bi$_2$Sr$_2$CuO$_6$, *Phys. Rev. B* **77**, 104506 (7pp) (2008).

[15] W. L. McMillan, Transition temperature of strong-coupled superconductors, *Phys. Rev.* **167**, 332-344 (1968).

[16] P. B. Allen and R. C. Dynes, Transition temperature of strong-coupled superconductors reanalyzed, *Phys. Rev. B* **12**(3), 905-922 (1975).

[17] D. Duan, Y. Liu, F. Tian, D. Li1, X. Huang, Z. Zhao, H. Yu, B. Liu, W. Tian, and T. Cui, Pressure-induced metallization of dense (H$_2$S)$_2$H$_2$ with high-Tc superconductivity, *Sci. Rep.* **4**, 1-6 (2014).

[18] P. Giannozzi, O. Andreussi, T. Brumme, O. Bunau, M. B. Nardelli, M. Calandra, R. Car, C. Cavazzoni, D. Ceresoli, M. Cococcioni, N. Colonna, I. Carnimeo, A. Dal Corso, S. de Gironcoli, P. Delugas, R. A. DiStasio Jr., A. Ferretti, A. Floris, G. Fratesi, G. Fugallo, R. Gebauer, U. Gerstmann, F. Giustino, T. Gorni, J. Jia, M. Kawamura, H.-Y. Ko, A. Kokalj, E. Küçükbenli, M. Lazzeri, M. Marsili, N. Marzari, F. Mauri, N. L. Nguyen, H.-V. Nguyen, A. Otero-de-la-Roza, L. Paulatto, S. Poncé, D. Rocca, R. Sabatini, B. Santra, M. Schlipf, A. P. Seitsonen, A. Smogunov, I. Timrov, T. Thonhauser, P. Umari, N. Vast, X. Wu, and S. Baroni, Advanced capabilities for materials modelling with Quantum ESPRESSO, *J. Phys.: Condens. Matter* **29**, 465901 (30pp) (2017).

[19] A. B. Migdal, Interaction between electrons and lattice vibrations in a normal metal, *Soviet Phys. JETP* **34**, 996-1001 (1958).

[20] Y. Nambu, Quasi-particles and gauge invariance in the theory of superconductivity, *Phys. Rev.* **117**, 648-663 (1960).

[21] R. Heid, Electron-phonon coupling, in the physics of correlated insulators, metals, and superconductors, edited by E. Pavarini, E. Koch, R. Scalettar, and R. Martin (Verlag, Jülich, 2017) Chapter 15.

[22] F. Marsiglio, Eliashberg theory: a short review, *Ann. Phys.* **417**, 1-23 (2020).

[23] E. N. Economou, Green's functions in quantum physics (Springer-Verlag, Berlin, 2006) p. 112.

[24] E. R. Margine and F. Giustino, Anisotropic Migdal-Eliashberg theory using Wannier functions, *Phys. Rev. B* **87**, 024505 (12pp) (2013).

[25] G. A. C. Ummarino, Eliashberg Theory in *Emergent phenomena in correlated matter,* Edited by E. Pavarini, E. Koch and U. Schollwöck (Verlag, Jülich, 2013) Chapter 13.

[26] S. J. Clark, M. D. Segall, C. J. Pickard, P. J. Hasnip, M. I. J. Probert, K. Refson, and M. C. Payne, First principles methods using CASTEP, *Z. Kristallogr.* **220**, 567-570 (2005).

[27] J. P. Perdew, K. Burke, and M. Ernzerhof, Generalized gradient approximation made simple, *Phys. Rev. Lett.* **77**, 3865-3868 (1996).

[28] D. H. Hamann, M. Schlüter, and C. Chiang, Norm-conserving pseudopotentials, *Phys. Rev. Lett.* **43**, 1494-1497 (1979).

[29] P. E. Blöchl, Projector augmented-wave method, *Phys. Rev. B* **50**, 17953-17979 (1994).

[30] A. Nojima, K. Yamashita, and B. Hellsing, Model Eliashberg functions for surface states, *Appl. Surf. Sci.* **254**, 7938-7941 (2008).

[31] H. Wang, J. S. Tse, K. Tanaka, T. Iitaka, and Y. Ma, Superconductive sodalite-like clathrate calcium hydride at high pressures, *Proc. Natl. Acad. Sci. U.S.A.* **109**, 6463–6466 (2012).

[32] H. Jeon, C. Wang, S. Liu, J. M. Bok, Y. Bang, and J.-H. Cho, Electron–phonon coupling and superconductivity in an alkaline earth hydride CaH$_6$ at high pressures, *New J. Phys.* **24**, 083048 (9pp) (2022).

[33] A. A. Darussalam and T. Koretsune, Superconductivity in CaH$_6$ and ThH$_{10}$ through fully ab initio Eliashberg method and self-consistent Green's function, *Sci. Rep.* **14**, 18399 (10pp) (2024).

[34] T. Wang, T. Nomoto, T. Koretsune, and R. Arita, Importance of self-consistency in first-principles Eliashberg calculation for superconducting transition temperature, *J. Phys. Chem. Solids* **178**, 111348 (5pp) (2023).